\documentstyle[12pt]{article}
\newcommand{\be}{\begin{equation}}
\newcommand{\ee}{\end{equation}}
\newcommand{\bea}{\begin{eqnarray}}
\newcommand{\eea}{\end{eqnarray}}
\newcommand{\vac}{\left.\mid 0\right>}
\newcommand{\k}{\vec{k}}
\newcommand{\q}{\vec{q}}
\newcommand{\s}{\vec{s}}
\newcommand{\x}{\vec{x}}
\newcommand{\y}{\vec{y}}
\newcommand{\p}{\vec{P}}
\newcommand{\R}{\vec{R}}
\newcommand{\r}{\vec{r}}
\newcommand{\tr}{\bigtriangleup}

\newcommand{\A}{\tilde{A}}
\newcommand{\op}{\hat{\vec{p}}}
\begin{document}

\begin{center}
{\Large \bf Bound states in nonrelativistic \\
four-fermion interaction model}.
\footnote{This work is partially supported by RFFR N 94-02-05204
and by grant "Universities of Russia", (St-Pb)
N 94-6.7-2057.} \\
{\sl A.V.Sinitskaya, E.V.Pavlova}. \\
{ \sl  Department of Theoretical Physics,
                 Irkutsk State University. \\
                 Irkutsk, 664003, RUSSIA.}\\
                 e-mail: VALL@physdep.irkutsk.su \\
\end{center}
\vspace{4mm}
\begin{abstract}
{\small The bound states of two particles are studied in frames of
non-relativistic quantum field model with current $\times$ current
type interaction by analyzing the Bethe-Salpeter amplitudes.
The Bethe-Salpeter equations are obtained in closed form.
The existence of Goldstone mode corresponding to the
spontaneous breaking of additional  SU(2) symmetry of the model is revealed.}
\end{abstract}
\vspace{4mm}

The conventional approach to the investigation of two particle
bound states in quantum field theory is based on obtaining the
Bethe-Salpeter equation (approximate in general) \cite{um}.
In the paper \cite{VLKS95} for the nonrelativistic model with current
$\times$ current type interaction two fermion bound states were examined
by straightforward solving of the two particle eigenstate problem for
the total Hamiltonian. Since the Heisenberg fields of this model
as in relativistic case contain both creation and annihilation operators
it seems instructive to follow up the obtaining of the Bethe-Salpeter
equation in the frames of this model.

In present paper it is shown that the existence of the closed Bethe-Salpeter
equation in this case is related with the fact that Hamiltonian
does not contain the "fluctuation" terms \cite{Schweb}, i.e.
 the terms which do not commute with the particle number operator.
\vspace{1cm}

Hamiltonian of the model has a form:
 \be
 H = \int d^3x \left[\Psi^{\dagger a}_\alpha (x)\varepsilon (\op)
    \Psi^a_\alpha(x)
   - \lambda J^\mu(x)J_\mu(x)\right] , \label{h}
 \ee
where $x = (\x,t)$,
\bea
&&\varepsilon (\op )e^{i\k\x} = \varepsilon (\k )e^{i\k\x },
\;\;\op = - i\vec{\nabla },\;\;
J^0 = \Psi^{\dagger a}_\alpha (x) \Psi^a_\alpha (x), \nonumber \\
&&\vec{J}(x) = \frac{1}{2mc}\left(\Psi^{\dagger a}_\alpha(x)\,\op
\Psi^a_\alpha(x) -
\,\op\Psi^{\dagger a}_\alpha(x) \Psi^a_\alpha(x)\right),
\eea
 $\alpha = 1,2$ is isospin index,
$a = 1,2$ is an index of the additional degrees of freedom.
$\varepsilon (\k) = \frac{\k^2}{2m} + mc^2$ - is a "bare" fermion spectrum.
$\Psi^a_\alpha (x)$ is a Heisenberg fermionic field which at $t = 0$
has the form:
\be
\Psi^a_\alpha (\x,0) =
\frac{1}{(2\pi)^{\frac{3}{2}}} \int d^3k  \{ e^{i\k\x} f^a (\k)
A_\alpha (\k) +  e^{- i\k\x} g^a (\k) \A^{\dagger}_\alpha (\k) \},
\label{psi}
\ee
where $A_\alpha (\k),\;\; \A_\alpha (\k)$ are annihilation operators of
two different kinds of fermions.
They satisfy the canonical anticommutation relations:
\be
\left\{ A_\alpha (\k),\; A^{\dagger}_\beta (\q)\right\} =
\left\{\A_\alpha (\k), \;\A^{\dagger}_\beta (\q)\right\} =
\delta_{\alpha\beta} \delta^3 (\k - \q),
\label{car}
\ee
The vacuum state is defined with respect to both kinds of particles
$A_\alpha (\k)$, $\A_\alpha (\k)$:
\be
A_\alpha\vac_{A\tilde{A}} = \tilde{A}_\alpha\vac_{A\tilde{A}} =  0
\label{vac}
\ee

The amplitudes $f^a (\k),\; g^a (\k)$ satisfy the completeness condition:
\be
f^a (\k) \bar{f}^b (\k) + g^a (- \k) \bar{g}^b (- \k)  =
\delta^{ab}; \quad a,b = 1,2.\label{ort}
\ee
It was shown at \cite{lev} that the existence of the exact
solution for the Hamiltonian implies the condition $f^{a},\;
g^{a}\,=\, const$.
At this condition Hamiltonian does not depend on $f^{a},\;g^{a}$,
what allows to take them in the following form:
\be
f^{a} = \left( \begin{array}{c}
             1 \\
               0 \end{array}  \right) = \delta_{a 1} ,\;\;
g^{a} = \left( \begin{array}{c}
             0 \\
             1 \end{array}  \right) = \delta_{a 2}.
\label{co}
\ee
So the field $\Psi^a_\alpha (x)$ can be presented via
"frequency" parts $\Psi^{(-)}_\alpha(x)$ and $\Psi^{(+)}_\alpha(x)$:
\be
\Psi^a_\alpha (x) =
f^a \Psi^{(-)}_\alpha(x)
+ g^a \Psi^{(+)}_\alpha(x),
\label{fpsi}
\ee
where
\bea
\Psi^{(-)}_\alpha(\x, 0) &=&
\frac{1}{(2\pi)^{\frac{3}{2}}} \int d^3k  e^{i\k\x}
 A_\alpha (\k) \nonumber \\
\Psi^{(+)}_\alpha(\x, 0) &=&
\frac{1}{(2\pi)^{\frac{3}{2}}} \int d^3k  e^{-i\k\x}
 \A^{\dagger}_\alpha (\k). \label{freq}
\eea

Let us write the Heisenberg equations for "frequency parts":
\bea
i\frac{\partial}{\partial t}\Psi^{(-)}_\alpha (x) =
& &[\Psi^{(-)}_\alpha (x), H]
= E_A\Psi^{(-)}_\alpha (x) + V^{\dagger}_\alpha (x),\nonumber\\
-i\frac{\partial}{\partial t}\Psi^{(+)}_\alpha (x) =
&-& [\Psi^{(+)}_\alpha (x), H] = E_{\A} \Psi^{(+)}_\alpha (x) +
\tilde{V}^{\dagger}_\alpha (x).\label{hei}
\eea

Because of the nonrenormalizability of the model under the consideration
one should introduce an ultraviolet cut-off $\Lambda$:
\bea
\frac{1}{V^*}\equiv \frac{1}{(2\pi)^3}\int \limits^\Lambda d^3k=
\frac{1}{6\pi^2}\Lambda^3; \quad
<k^2> \equiv \frac{\int \limits^\Lambda \k^2 d^3k}{\int\limits^\Lambda d^3k}
=\frac{3}{5}\Lambda^2;
\;\;\;g \equiv \frac{\lambda}{V^*}.
\label{g}
\eea
Renormalized coupling constant $g$ has dimension of energy and enters
alone into the final expressions for the all dynamical characteristics.

There are two point of views on the value of cut-off
parameter $\Lambda$. First one is to remain $\Lambda$ finite
choosing it by physical sense (\cite{klev},\cite{kalin}). The second
point of view is to consider $\Lambda$ as a regularization parameter
only and tend it to infinity in final expressions, supposing the definite
behavior over $\Lambda$ of "bare" quantities (\cite{fad}).

Direct calculations in (\ref{hei}) lead to the
expressions for one particle energies:
\bea
&& E_A(\k) = \frac{k^2}{2m_A} + mc^2 -5g
          + g\frac{<k^2>}{4m^2c^2};
\nonumber \\
&& E_{\A} (\k) = \frac{k^2}{2m_{\tilde{A}}} - mc^2 + 3g
+ g\frac{<k^2>}{4m^2c^2};
\label{spectra} \\
&& m_A = \frac{m}{\frac{g}{2mc^2} + 1};\;\;\;
m_{\A} = \frac{m}{\frac{g}{2mc^2} - 1};\;\;\;
m = \frac{m_A}{2}\left(1 + \sqrt{1 + \frac{2g}{m_Ac^2}}\right).
\nonumber
\eea
\bea
&&V^{\dagger}_\alpha (x) = - 2 \lambda \Psi^{\dagger (+)}_\gamma (x)
\Psi^{(-)}_\gamma (x)\Psi^{(-)}_\alpha (x)
+ 2 \lambda \Psi^{(+)}_\gamma (x)
\Psi^{\dagger (-)}_\gamma (x)\Psi^{(-)}_\alpha (x) + \nonumber \\
&&+ \frac{4 \lambda}{(2mc)^2}\Psi^{\dagger (+)}_\gamma (x)
\op \Psi^{(-)}_\gamma (x)\op \Psi^{(-)}_\alpha (x)
- \frac{4 \lambda}{(2mc)^2}\op\Psi^{\dagger (+)}_\gamma (x)
 \Psi^{(-)}_\gamma (x)\op \Psi^{(-)}_\alpha (x) + \nonumber \\
&&+ \frac{4 \lambda}{(2mc)^2}\Psi^{(+)}_\gamma (x)
\op \Psi^{\dagger (-)}_\gamma (x)\op \Psi^{(-)}_\alpha (x)
- \frac{4 \lambda}{(2mc)^2}\op \Psi^{(+)}_\gamma (x)
\Psi^{\dagger (-)}_\gamma (x)\op \Psi^{(-)}_\alpha (x) + \nonumber \\
&&+ \frac{2 \lambda}{(2mc)^2}\op\left(\Psi^{\dagger (+)}_\gamma (x)
\op \Psi^{(-)}_\gamma (x)\right) \Psi^{(-)}_\alpha (x)
- \frac{2 \lambda}{(2mc)^2}\op\left(\op\Psi^{\dagger (+)}_\gamma (x)
 \Psi^{(-)}_\gamma (x)\right) \Psi^{(-)}_\alpha (x) + \nonumber \\
&&+ \frac{2 \lambda}{(2mc)^2}\op\left(\Psi^{(+)}_\gamma (x)
\op \Psi^{\dagger (-)}_\gamma (x)\right) \Psi^{(-)}_\alpha (x)
- \frac{2 \lambda}{(2mc)^2}\op\left(\op\Psi^{(+)}_\gamma (x)
 \Psi^{\dagger (-)}_\gamma (x)\right) \Psi^{(-)}_\alpha (x), \nonumber \\
&&\tilde{V}^{\dagger}_\alpha (x) = 2 \lambda \Psi^{\dagger (+)}_\gamma (x)
\Psi^{(-)}_\gamma (x)\Psi^{\dagger (-)}_\alpha (x)
- 2 \lambda \Psi^{(+)}_\gamma (x)
\Psi^{\dagger (-)}_\gamma (x)\Psi^{\dagger (-)}_\alpha (x) + \label{VE} \\
&&+ \frac{4 \lambda}{(2mc)^2}\Psi^{\dagger (+)}_\gamma (x)
\op \Psi^{(-)}_\gamma (x)\op \Psi^{\dagger (-)}_\alpha (x)
- \frac{4 \lambda}{(2mc)^2}\op\Psi^{\dagger (+)}_\gamma (x)
 \Psi^{(-)}_\gamma (x)\op \Psi^{\dagger (-)}_\alpha (x) + \nonumber \\
&&+ \frac{4 \lambda}{(2mc)^2}\Psi^{(+)}_\gamma (x)
\op \Psi^{\dagger (-)}_\gamma (x)\op \Psi^{\dagger (-)}_\alpha (x)
- \frac{4 \lambda}{(2mc)^2}\op \Psi^{(+)}_\gamma (x)
\Psi^{\dagger (-)}_\gamma (x)\op \Psi^{\dagger (-)}_\alpha (x) + \nonumber \\
&&+ \frac{2 \lambda}{(2mc)^2}\op\left(\Psi^{\dagger (+)}_\gamma (x)
\op \Psi^{(-)}_\gamma (x)\right) \Psi^{\dagger (-)}_\alpha (x)
- \frac{2 \lambda}{(2mc)^2}\op\left(\op\Psi^{\dagger (+)}_\gamma (x)
 \Psi^{(-)}_\gamma (x)\right) \Psi^{\dagger (-)}_\alpha (x) + \nonumber \\
&&+ \frac{2 \lambda}{(2mc)^2}\op\left(\Psi^{(+)}_\gamma (x)
\op \Psi^{\dagger (-)}_\gamma (x)\right) \Psi^{\dagger (-)}_\alpha (x)
- \frac{2 \lambda}{(2mc)^2}\op\left(\op\Psi^{(+)}_\gamma (x)
 \Psi^{\dagger (-)}_\gamma (x)\right) \Psi^{\dagger (-)}_\alpha (x).
\nonumber
\eea
As one can see threelinear terms
$V^{\dagger}_\alpha, \tilde{V}^{\dagger}_\alpha$ acting on vacuum vanish
\be
V^{\dagger}_\alpha\vac = \tilde{V}^{\dagger}_\alpha\vac = \left<0
\right.\mid V^{\dagger}_\alpha = \left<0 \right.\mid\tilde{V}^{\dagger}_\alpha
= 0,
\ee
due to the absence of "fluctuational terms" in the Hamiltonian (\ref{h}).
This fact turns out to be essential condition for Bethe-Salpeter equation
to have closed form.

Let us consider the Bethe-Salpeter (BS) amplitudes for fermionic fields
$\Psi^a_\alpha (x)$:
\bea
&& G^{AA, ab}_{\alpha\beta} (x,y;\p) = \left<0 \right.\mid T\left(\Psi^a
_\alpha (x)\Psi^b_\beta (y)\right)\left.\mid D_{AA} (\p) \right>,
\nonumber\\
&& G^{\A\A ,ab}_{\alpha\beta} (x,y;\p) = \left<0 \right.\mid T
\left(\Psi^{\dagger a}
_\alpha (x)\Psi^{\dagger b}_\beta (y)\right)\left.\mid D_{\A\A} (\p) \right>,
\nonumber\\
&& G^{A\A, ab}_{\alpha\beta} (x,y;\p) = \left<0 \right.\mid T\left(\Psi^a
_\alpha (x)\Psi^{\dagger b}_\beta (y)\right)\left.\mid D_{A\A} (\p) \right>,
\label{BS}
\eea
where $\left.\mid D_{AA} (\p) \right>,\; \left.\mid D_{\A\A} (\p) \right>$
and $\left.\mid D_{A\A} (\p) \right>$ are two particle bound states
of three possible kinds: $A - A$, $\A - \A$, $A - \A$
particles respectively. The action of Hamiltonian on these states gives
the equation for bound states energies $\mu_{AA} (\p),\;\; \mu_{\A\A} (\p),
\;\;\mu_{A\A} (\p)$:
\bea
H \left.\mid D_{AA} (\p)\right> =
(W_0 + \mu_{AA}(\vec{P}))\left.\mid D_{AA} (\p)\right>,\nonumber \\
H \left.\mid D_{\A\A} (\p)\right> =
(W_0 + \mu_{\A\A}(\vec{P}))\left.\mid D_{\A\A} (\p)\right>, \nonumber \\
H \left.\mid D_{A\A} (\p)\right> =
(W_0 + \mu_{A\A}(\vec{P}))\left.\mid D_{A\A} (\p)\right>.
\eea
 $W_0$ is the vacuum energy.

Let us rewrite the BS amplitudes via "frequency" parts. One can see
that matrix elements containing creation operators vanish. Therefore,
we have:
\bea
G^{AA, ab}_{\alpha\beta} (x,y;\p) = f^af^b\left<0 \right.\mid T
\left(\Psi^{(-)}_\alpha (x)\Psi^{(-)}_\beta (y)\right)
\left.\mid D_{AA} (\p) \right>, \nonumber \\
G^{\A\A, ab}_{\alpha\beta} (x,y;\p) = \bar{g}^a\bar{g}^b\left<0 \right.\mid T
\left(\Psi^{\dagger (-)}_\alpha (x)\Psi^{\dagger (-)}_\beta (y)\right)
\left.\mid D_{\A\A} (\p) \right>,  \nonumber \\
G^{A\A, ab}_{\alpha\beta} (x,y;\p) = f^a\bar{g}^b\left<0 \right.\mid T
\left(\Psi^{(-)}_\alpha (x)\Psi^{\dagger (-)}_\beta (y)\right)
\left.\mid D_{A\A} (\p) \right>.
\eea
So instead of BS amplitudes (\ref{BS}) we will consider the ones written
via "frequency" parts which carry the isospin indexes only:
\bea
G^{AA, ab}_{\alpha\beta} (x,y;\p) = f^af^b G^{AA}_{\alpha\beta} (x,y;\p) ,
\nonumber \\
G^{\A\A, ab}_{\alpha\beta} (x,y;\p) =
\bar{g}^a\bar{g}^b G^{\A\A}_{\alpha\beta} (x,y;\p) , \nonumber \\
G^{A\A, ab}_{\alpha\beta} (x,y;\p) =
f^a\bar{g}^b G^{A\A}_{\alpha\beta} (x,y;\p) .
\eea
Thus using (\ref{hei}) we derive differential  equations on these BS
amplitudes. Let us write for example the one for
$G^{AA}_{\alpha\beta} (x,y;\p)$:
\bea
&& \left(i\frac{\partial}{\partial t_x}
- E_A (\nabla_x)\right)
G^{AA}_{\alpha\beta} (x,y;\p) =
\left<0 \right.\mid T\left(V^{\dagger}_\alpha (x)
\Psi^{(-)}_\beta (y)\right)\left.\mid D_{AA} (\p) \right>,
\nonumber \\
&& \left(i\frac{\partial}{\partial t_y} - E_A (\nabla_y)\right)
\left(i\frac{\partial}{\partial t_x} - E_A (\nabla_x)\right)
G^{AA}_{\alpha\beta} (x,y;\p) =
\nonumber \\
&& = - i \delta (t_x - t_y)
\left<0 \mid \left\{V^{\dagger}_\alpha (x),
\Psi^{(-)}_\beta (y)\right\}\mid D_{AA} (\p) \right> =
\label{Umez} \\
&&= - 2 i \lambda\left\{1-\frac{1}{(2mc)^2}\left[\op^{\,2}_{\xi} -
\op^{\,2}_x
+ 2 \op_{\eta}(\op_{\xi} - \op_x)\right]\right\}\delta^4 (x - y)
G^{AA}_{\alpha\beta} (\eta,\xi;\p)\big|_{x = \xi = \eta}.
\nonumber
\eea
Here it is taken into account that at equal times
\[
\left<0 \right.\mid T\left(\Psi^{(-)}_\alpha (x)
\Psi^{(-)}_\beta (y)\right)\left.\mid  D_{AA} (\p) \right>
\big|_{t_x = t_y = t} = \left<0 \right.\mid \Psi^{(-)}_\alpha (\x,t)
\Psi^{(-)}_\beta (\y,t)\left.\mid D_{AA} (\p) \right>
\big|_{t_x = t_y = t}.
\]
The corresponding expressions for $G^{\A\A}_{\alpha\beta} (x,y;\p)
\mbox{ and }G^{A\A}_{\alpha\beta} (x,y;\p)$ could be
obtained by the same way. The first of them has the similar form as
(\ref{Umez}), and the last one differs by sign of contribution from
zero component $J^0 (x)$ in the Hamiltonian (\ref{h}).

The expression (\ref{Umez}) can be rewritten in the integral form:
\bea
&&G^{AA}_{\alpha\beta} (x,y;\p) =
2 i \lambda \int d^4 \sigma d^4 z \tr (x - z)
\tr (y - \sigma)
\left\{ 1 - \right. \\
&&- \left.\frac{1}{(2mc)^2}\left[\op^{ 2}_{\xi} - \op^{ 2}_x
+ 2 \op_{\eta}(\op_{\xi} - \op_x)\right]\right\}
\delta^4 (z - \sigma)
G^{AA}_{\alpha\beta} (\eta,\xi;\p)\big|_{z = \xi = \eta},
\nonumber
\eea
where $\tr (x)$ is the casual (coinciding with retarded) Green function,
satisfying the equation:
\bea
\left( i \frac{\partial}{\partial t_x} - E_A (\nabla_x) \right)
\tr (x - z) = i \delta^4 (x -z).
\eea

Putting now $t_x = t_y = t$ we obtain the equation for instantaneous
BS amplitude:
\bea
&&G^{AA}_{\alpha\beta} (\x,\y,t;\p) =
2 i \lambda \int d^4 z \bigtriangleup (x - z)
\left\{ 1 - \right.
\label{intBS} \\
&&- \left.\frac{1}{(2mc)^2}\left[\op^{ 2}_{\xi} - \op^{ 2}_x
+ 2 \op_{\eta}(\op_{\xi} - \op_x)\right]\right\}
\bigtriangleup (y - z)
G^{AA}_{\alpha\beta} (\eta,\xi;\p)\big|_{z = \xi = \eta}.
\nonumber
\eea
Let us pass to new variables
$\R = \frac{1}{2} (\x + \y),\; \r = \x - \y $
- total and relative coordinates respectively. Then BS amplitude
has the form:
\be
G^{AA}_{\alpha\beta} (\x,\y,t;\p) =
e^{- i\mu_{AA}(\p)t + i\p\R} K^{AA}_{\alpha\beta} (\r,0;\p),
\ee
where
\be
K^{AA}_{\alpha\beta} (\r,0;\p) =
\left<0 \right.\mid \Psi^{(-)}
_\alpha (\frac{\r}{2},0)\Psi^{(-)}_\beta (\frac{- \r}{2},0)\left.
\mid D_{AA} (\p) \right>.
\label{J}
\ee
Combaining all mentioned above and passing  in the eq (\ref{intBS})
to the momentum representation (the expressions for another two
cases are obtained by the same way) we have:
\bea
F^{AA,\A\A}_{\alpha\beta}(\s; \p) =
\lambda \frac{2}{(2\pi)^3}\int d^3k
\frac{1 + (2mc)^{-2}[(\k + \s)^2 - \p^2]}
{E_{A,\A}(\frac{\p}{2} + \k) + E_{A,\A}(\frac{\p}{2} - \k) -\mu_{AA,\A\A} (\p)}
F^{AA,\A\A}_{\alpha\beta}(\k; \p), \label{F-AA} \\
F^{A\A}_{\alpha\beta}(\s; \p) =
\lambda \frac{2}{(2\pi)^3}\int d^3k
\frac{-1 + (2mc)^{-2}[(\k + \s)^2 - \p^2]}
{E_A(\frac{\p}{2} + \k) + E_{\A}(\frac{\p}{2} - \k) -\mu_{A\A} (\p)}
F^{A\A}_{\alpha\beta}(\k; \p).  \label{F-A-A}
\eea
This is exactly the equations obtained in frameworks of eigenstate problem.
The functions
\bea
D^{AA}_{\alpha\beta}(\s; \p) =
\frac{F^{AA}_{\alpha\beta}(\s; \p)}
{\left(E_A(\frac{\p}{2} + \s) + E_A(\frac{\p}{2} - \s) -\mu_{AA} (\p)
\right)}, \label{uraV}\\
D^{\A\A}_{\alpha\beta}(\s; \p) =
\frac{F^{\A\A}_{\alpha\beta}(\s; \p)}
{\left(E_{\A}(\frac{\p}{2} + \s) + E_{\A}(\frac{\p}{2} - \s) -\mu_{\A\A} (\p)
\right)}, \nonumber\\
D^{A\A}_{\alpha\beta}(\s; \p) =
\frac{F^{A\A}_{\alpha\beta}(\s; \p)}
{\left(E_A(\frac{\p}{2} + \s) + E_{\A}(\frac{\p}{2} - \s) -\mu_{A\A} (\p)
\right)}, \nonumber
\eea
have the meaning of bound states wave functions.

So we have got the closed equations for
$D^{AA}_{\alpha\beta} (\k;\p),\;\; D^{\A\A}_{\alpha\beta} (\k;\p)$
and $D^{A\A}_{\alpha\beta} (\k;\p)$. The solutions for (\ref{F-AA},\ref{F-A-A})
would have the following form:
\be
F_{\alpha\beta} = A_{\alpha\beta} (\p) + \k^2 B_{\alpha\beta} (\p)
+ \k \vec{C}_{\alpha\beta} (\p).
\label{solut}
\ee
Here $A_{\alpha\beta} (\p),\;\; B_{\alpha\beta} (\p),\;\;
\vec{C}_{\alpha\beta} (\p)$ are matrixes depending only on $\p$.

\vspace{.4cm}

I). First, let us consider  the equations (\ref{F-AA}) for the bound states
of the same kind particles, analyzing the bound state rest-frame case,
that is $\p = 0$. As one can see, the wave functions for these bound
states have a definite symmetry:
\be
D^{AA}_{\alpha\beta} (\k; 0) = - D^{AA}_{\beta\alpha} (- \k; 0),\;\;\;
D^{\A\A}_{\alpha\beta} (\k; 0) = - D^{\A\A}_{\beta\alpha} (- \k; 0).
\label{simchet}
\ee
Therefore symmetrical and skewsymmetrical parts of $D^{AA} (\k; 0),\;
D^{\A\A} (\k; 0)$ are splitted. Then in solution (\ref{solut})
($A_{\alpha\beta}, B_{\alpha\beta}, \vec{C}_{\alpha\beta}$ are constant
matrices) skewsymmetrical $A_{\alpha\beta}$ and $B_{\alpha\beta}$ and
symmetrical $\vec{C}_{\alpha\beta}$ over $\alpha,\; \beta$ matrices contribute
independently to the bound states and correspond to isoscalar and isovector
states.
Hence it follows, that $A_{\alpha\beta} = A\epsilon_{\alpha\beta}, \;\;\;
B_{\alpha\beta} =
B\epsilon_{\alpha\beta}$ and $\vec{C}_{\alpha\beta}$ can be expanded
over three symmetrical matrices: $I,\;\tau_1,\;\;\tau_3$.
According to these remarks
the equation (\ref{F-AA}) is brought to the following set of equations:
\bea
&&A = I_0A + (2mc)^{-2}I_1A + I_1B + (2mc)^{-2}I_2B, \nonumber \\
&&B = (2mc)^{-2}I_0A + (2mc)^{-2}I_1B, \nonumber \\
&&C^i_{\alpha\beta} = (2mc)^{-2}I^{ij} C^j_{\alpha\beta}.
\label{sys}
\eea
Here
\bea
I_n = \frac{2 \lambda}{(2\pi)^3}\int d^3k\frac{(\k^2)^n}{2E_A (\k) - \mu_s};
\;\;\;
I^{ij} = \frac{2 \lambda}{(2\pi)^3}\int d^3k
\frac{2\k^i\k^j}{2E_A (\k) - \mu_v}.
\label{I's}
\eea
$\mu_s, \;\mu_v \equiv \mu_{AA}(0), \;\mu_{\A\A}$ stand for isoscalar
and isovector states masses of bound states $AA, \;\A\A$ accordingly.
Let us introduce the dimensionless parameter $G = 2g/(m_Ac^2)$
and the parameters:
\be
\chi^2_s = m_A(2m_Ac^2 - \mu_s), \;\;\;
\chi^2_v = m_A(2m_Ac^2 - \mu_v). \label{Xa}
\ee
The equations on these parameters $\chi_s,\;\;\chi_v$ follow from (\ref{sys}).

From the last relation of (\ref{sys}) using (\ref{I's}) we obtain usual
"gap" equation for isovector state:
\be
1 = \frac{\lambda}{(2\pi)^3}\frac{m_A}{3m^2c^2}\int \limits^\Lambda
 d^3k \frac{\k^2}{\k^2 + \chi^2_v}\label{mu-v}
\ee

First two equations (\ref{sys}) form linear homogeneous system with
respect to $A$ and $B$. Demanding the determinant of this system to be zero
we come to the equation on
$\mu_s$:
\be
\left(I_1 - (2mc)^2\right)^2 = I_0\left(I_2 + (2mc)^4\right).
\label{det}
\ee
Then from (\ref{det}) the equation on $\chi^2_s$ follows:
\bea
\frac{\lambda m_A}{(2\pi)^3}\int \limits^\Lambda \frac{d^3k}{\k^2 +
\chi^2} =
\left(\frac{m_A g}{(2mc)^2} - \frac{1}{2}\right)^2\left[\frac{1}{2} -
\frac{\chi^2}{(2mc)^2} + \frac{m_A g}{(2mc)^2}\cdot
\frac{<\k^2> + \chi^2}{(2mc)^2}\right]^{-1}
\label{eq3}
\eea
As it was shown  in \cite{lev} there are three inequivalent operator
realizations of the Hamiltonian (\ref{h}) called "A", "B" and "C" cases.
The demand in  "B" case corresponding to zero vacuum energy for the
one particle energy to have the usual nonrelativistic form,
$ E_B(\k)=(2m_A)^{-1}k^2+m_Ac^2$, leads to the expressions:
$<k^2> = m_A^2c^2 \left[1 + G + \sqrt{1 + G}\right]$,
allowing to exclude the parameter $<k^2>$. Then
after integration in the l.h.s. of equation (\ref{eq3})
we derive transcendental equation:
\be
\bigl(z^2c_1 - c_2\bigr)\bigl(z - \arctan z\bigr) = z^3, \label{tan}
\ee
where
$$
c_1 = \frac{9}{20}\cdot\frac{3 + 2G + \sqrt{1 + G}}{1 + G + \sqrt{1 + G}},
\;\;\;c_2 = \frac{3}{4}G\biggl(1 + \frac{2}{1 + \sqrt{1 + G}}\biggr),\;\;\;
z = \frac{\Lambda}{\chi_s}
$$
The analysis of this equation shows that it always has a
solution at $c_1 > 1$. With a good accuracy  $c_1 \simeq \frac{9}{10}G$,
therefore, the condition of the existence of a solution
for bound state with zero isospin reduces to
$G > \frac{10}{9}$. From the numerical
solution of the equation (\ref{tan}) follows that $z(G)$
strongly changes in the region $\frac{10}{9} < G \leq 1.3$, but
further on, at $G\geq 1.3$, is slowly achieving its asymptotic  value
$z(\infty) = \sqrt{\frac{5}{6}}$.

\vspace{.4cm}

II).Now let us consider the equation (\ref{F-A-A}) for wave function of
the two different kind particle bound state. In this case there is no
such a symmetry as took place for bound states of particles $AA
\mbox{ and }\A\A$ (\ref{simchet}). Therefore the solution (\ref{solut})
can not be splitted, and one should solve the homogeneous system of three
equations for matrixes $A_{\alpha\beta} (\p)$, $B_{\alpha\beta}
(\p)$, $C^i_{\alpha\beta} (\p)$. The denominator of (\ref{F-A-A}) contains
two different one-particle excitations spectra $E_A (\frac{\p}{2} + \k)$
and $E_{\A} (\frac{\p}{2} - \k)$. Using the explicit form of these
spectra (\ref{spectra}) and changing the variable $\k\;\rightarrow \;\;
\vec{\kappa}\; = \;\k \;+ \;\p\gamma $ where $ \gamma \;= \;
mc^2/g \;= \;(1 + \sqrt{1 + G})/G$
one could obtain the same (symmetrical over $\vec{\kappa}$) structure
of the denominator
in the kernel of the equation (\ref{F-A-A}) as in previous case.
The system of equations to determine the matrixes
$A_{\alpha\beta} (\p),\;\; B_{\alpha\beta}
(\p),\;\; C^i_{\alpha\beta} (\p)$ now reads:
\bea
&& \left( -(2mc)^2I_0 (\p) + \p^2 ((2\gamma)^2 - 1)I_0 (\p) + I_1 (\p) -1
\right)
A_{\alpha\beta} (\p) +  \nonumber \\
&& + \left( -(2mc)^2 + \p^2
((2\gamma)^2 - 1)I_1 (\p) +
I_2 (\p)\right)B_{\alpha\beta} (\p)
- \frac{4}{3}\gamma I_1 (\p) \p
\vec{C}_{\alpha\beta} (\p) = 0, \nonumber \\
&& I_0 (\p) A_{\alpha\beta} (\p) + (I_1 (\p) - 1) B_{\alpha\beta} (\p) = 0,
\label{syst} \\
&& 4\gamma I_0 (\p) \p A_{\alpha\beta} (\p) + 4\gamma I_1 (\p)
\p B_{\alpha\beta} (\p) -
\left(\frac{2}{3}I_1 (\p) - 1\right)\vec{C}_{\alpha\beta} (\p) = 0,
\nonumber \\
&& \mbox{where}\quad
M^2 (\p) = \frac{\p^2}{4}(1 - 4\gamma^2) - (2mc)^2 + <\k^2> - 2m\gamma
\mu (\p);
\nonumber \\
&& I_n (\p) = \frac{V^*}{(2\pi)^3}\int
\frac{d^3s\, (\s^{\,\,2})^n}{\s^{\,\,2} + M^2 (\p)},
\label{II}
\eea
As it is seen from the last equation of (\ref{syst}),
the vector $\vec{C}_{\alpha\beta} (\p)$ has the structure
$\vec{C}_{\alpha\beta} (\p) \; =\; \p C_{\alpha\beta} (\p)
\;+ \;\vec{C}^0_{\alpha\beta} (\p)$. Substituting this expression
into the system (\ref{syst}), we obtain at first the equation
\be
\frac{2}{3}I_1 (\p) \vec{C}^0_{\alpha\beta}= \vec{C}^0_{\alpha\beta},
\quad,
\label{ceo}
\ee
and secondly the condition for determinant of the system:
\bea
&&\left( -(2mc)^2I_0 (\p) + \p^2 ((2\gamma)^2 - 1)I_0 (\p) + I_1 (\p) -1 \right)
A_{\alpha\beta} (\p) +  \nonumber \\
&&+ \left( -(2mc)^2 + \p^2
((2\gamma)^2 - 1)I_1 (\p) +
I_2 (\p)\right)B_{\alpha\beta} (\p)
- \frac{4}{3}\gamma I_1 (\p) \p^2
C_{\alpha\beta} (\p) = 0, \nonumber \\
&&I_0 (\p) A_{\alpha\beta} (\p) + (I_1 (\p) - 1) B_{\alpha\beta} (\p) = 0,
\label{sysnew} \\
&&4\gamma I_0 (\p)  A_{\alpha\beta} (\p) + 4\gamma I_1 (\p)  B_{\alpha\beta} 
(\p) -
\left(\frac{2}{3}I_1 (\p) - 1\right)C_{\alpha\beta} (\p) = 0 \nonumber
\eea
to be equal to zero.
So the equation  (\ref{ceo}) determines the solution for isovector state:
\be
\frac{2}{3(2\pi)^3}\int
\frac{d^3s\, \s^{\,\,2}}{\s^{\,\,2} + M^2_1 (\p)} = \frac{1}{V^*}.
\ee
The system (\ref{sysnew}) has no fixed tensor structure so the
mentioned condition for determinant of the (\ref{sysnew})
gives the equation for bound state mass $\mu_{A\A} (\p)$
of isoscalar and isovector states in the similar form:
\be
\mu^s_{A\A} (\p) = \mu^{v,2}_{A\A} (\p) =  \frac{G\p^2}{4m_A}\left[5 +
12 \gamma^2 - \frac{24\gamma^2}{1 + 2M^2_2 (\p) I_0 (\p)}\right].
\ee
At infinitesimal $\p \mbox{ bound state energy }\mu_{A\A} (\p)$ tends to zero,
that corresponds to the Goldstone mode. The existence
of these four Goldstone modes was shown algebraically in the paper \cite{lev}.
Such a state is generated by the operator $Q^{\dagger}_{\alpha\beta} =
\int d^3 k A_\alpha^\dagger (\k) \tilde{A}_\alpha^\dagger (-\k) $:
\be
Q\vac = \left.\mid D_{A\A} (0)\right>.
\ee
Therefore the wave function $D^{A\A}_{\alpha\beta} (\k;0) = 1$ at $\p = 0$.
Let us substitute the solution (\ref{solut}) into eq.(\ref{F-A-A}) at
$\p = 0$, taking into account that $\vec{C}_{\alpha\beta} (0) = 0$.
Thus we obtain the relation
\be
-2g + g\frac{\k^2}{2m^{2}c^{2}} + g\frac{<\k^2>}{2m^{2}c^{2}} =
E_{A}(\k) + E_{\A}(-\k),
\ee
which according to (\ref{spectra}) is identity independently of the
cut-off $\Lambda$ value.

The authors are grateful to A.N.Vall, S.E.Korenblit and V.M.Leviant
for inspiring discussions.


\begin{thebibliography}{7}
\bibitem {um} H.\,Umezawa, H.\,Matsumoto, M.\,Tachiki,
Thermo Field Dynamics and Condensed States, North - Holland
Publishing Company 1982.
\bibitem {VLKS95} A.N.\,Vall,  S.E.\,Korenblit,  V.M.\,Leviant  and
A.V.\,Sinitskaya, Preprint ISU-IAP. Th95-02, Irkutsk 1995. quant-ph/9508017.
\bibitem{Schweb} Silvan S.\,Schweber, An Introductin to Relativistic
Quantum Field Theory, Petrson \& Co., 1961
\bibitem{lev} A.N.\,Vall,  S.E.\,Korenblit,  V.M.\,Leviant  and
A.V.\,Sinitskaya, Proceedings  of  X  International  WorkShop,
High Energy Physics and Quantum Field Theory (Zvenigorod 1995),
Moscow State University Publishing House (to be published).
\bibitem{klev} S.P.\,Klevansky, Reviews of Modern Physic,
Vol.{\bf 64}, No.3, July 1992.
\bibitem{kalin} V.N.\,Pervushin, Yu.L.\,Kalinovsky, W.\,Kallies,
N.A.\,Sarikov, Fortschr. Phys. {\bf 38} (1990) 5, 333-351.
\bibitem{fad} F.A.\,Beresin, L.D.\,Faddeev, Dokl.Akad.Nauk SSSR
{\bf 137} (1961) No.5

\end{thebibliography}
\end{document}